\documentclass[twocolumn,prl,aps,epsf,floats]{revtex4}

	\newif\ifpdf
	\ifx\pdfoutput\undefined
	\pdffalse 
	\else
	\pdfoutput=1 
	\pdftrue
	\fi
	\ifpdf
	\usepackage[pdftex]{graphicx}
	\else
	\usepackage{graphicx}
	\fi

\begin{document}
 \ifpdf
 \DeclareGraphicsExtensions{.pdf, .jpg, .tif}
 \else
 \DeclareGraphicsExtensions{.eps, .jpg}
 \fi


\title{Microwave resonances of the bubble phases in 1/4 and 3/4 
filled higher Landau levels}
\author{R.\ M.\ Lewis$^{1,2}$, P.\ D.\ Ye$^{1,2}$, L.\ W.\ 
Engel$^{1}$, 
D.\ C.\ Tsui$^{2}$, L.\ N.\ Pfeiffer$^{3}$, and K.\ W.\ West$^{3}$}

\address{$^{1}$National High Magnetic Field Lab and Department of Physics,
Florida State University, Tallahassee, FL 32306, USA\\
$^{2}$Department of Electrical Engineering, Princeton University, Princeton, NJ 08544\\
$^{3}$Bell Laboratories, Lucent Technology, Murray Hill, NJ 07974}

\date{\today}

\begin{abstract}
We have measured the diagonal conductivity, $\sigma_{xx}$, in the 
microwave regime of an ultrahigh mobility 
two dimensional electron system.  We find a sharp resonance in 
Re[$\sigma_{xx}$] versus frequency when $\nu>4$ and the 
partial filling of the highest Landau level, $\nu^*$,  is $\sim 1/4$ 
or  3/4 and temperatures $<0.1$ K.  The resonance 
appears for a range of $\nu^*$ from 0.20 to 0.37  and again from 0.62 
to 0.82.  The peak frequency, $f_{pk}$ changes from $\sim$500 to $\sim$150 MHz as 
$\nu^*=1/2$ is approached.  This range of $f_{pk}$ shows no dependence on $\nu$ where the resonance is observed. The quality factor, $Q$, of the resonance is maximum at about $\nu^*=0.25$ and 
0.74.  We interpret the resonance as due to a 
pinning mode of the bubble phase crystal. 

\pacs{:}

\end{abstract}
\maketitle    

Two dimensional electron systems, 2DES, confined in ultra clean GaAs/AlGaAs heterostructures and subjected to perpendicular magnetic fields, $B$, show strongly anisotropic diagonal resistance at half integer filling factors, $\nu=$ 9/2, 
11/2, 13/2, ...\cite{mlilly1,rrdu} for temperatures, $T< 150$ mK.  The same studies find that minima appear in the diagonal resistance at partial fillings of $\nu^*\sim 1/4$ and 3/4, where $\nu^*=\nu-$[$\nu$], $\nu>$ 4, and [$\nu$] is the greatest integer less than $\nu$.  Concomitantly the Hall resistance is quantized to the value of the adjacent 
integer quantum Hall effect plateau, and hence these states have been christened the re-entrant integer quantum Hall 
effect,  RIQHE\cite{cooper2}.  Previously constructed theories\cite{foglerkoulakov96,chalkerandmossner} 
made predictions with which these observations are consistent.  In particular, it was 
proposed that the RIQHE was due to an isotropic solid phase of the 2DES---a 
regular triangular crystal lattice with two or more electron guiding 
centers per lattice site.  This crystal, known as the ``bubble'' phase, would be
insulating because of pinning by disorder.

At this time, the case for viewing the RIQHE as a manifestation of 
the bubble phase within the context of the theory rests (i) on its location between the stripe 
phase and the integer quantum Hall effect plateau and (ii) on the 
observed insulating behavior.  In addition, one experimental study has 
seen non--linear I--V data in 1/4 and 3/4 filled levels\cite{cooper1}, a possible indication of a depinning transition.  These data are consistent with a crystalline phase, but many phenomena, e.\ g.\  ``breakdown"\cite{qhebreakdown}, can cause nonlinearities in transport.  Thus, whether or not the 2DES forms a bubble phase at these fillings is still in question.

The observation of solid phases of the 2DES in GaAs heterostructures 
has so far focused on the Wigner crystal, WC\cite{mshayegan}, at 
very high magnetic fields $\nu <$ 1/5.  Like the bubble phase, the WC 
is a triangular lattice of electrons 
pinned by disorder and therefore insulating at low temperatures.  
One technique used to study the WC is to excite the pinning mode\cite{fl1,normand} 
of the lattice with microwave radiation at about 1 GHz.  The pinning mode describes 
the oscillation of  domains of WC, with length scales of many lattice spacings, within 
the disorder potential of the sample.  These experiments\cite{lloydssc,ccliprl97,ccliprb}
 find a narrow resonance in the real part of the diagonal conductivity, Re[$\sigma_{xx}$],
 versus frequency, $f$.

	In this letter we present measurements of the microwave conductivity 
in the higher Landau levels, $\nu>$ 4, of an ultrahigh quality 2DES.  
Our data show a strong resonance in Re[$\sigma_{xx}$] versus $f$,  with
 $Q \sim 3$ around $\nu^{*}=1/4$ and 3/4  fillings
when $T \le 110$ mK and $Q$ is the peak frequency over the full width at half maximum.
The resonance is interpreted as a pinning mode.  It follows that the 
resonance is direct evidence that the 2DES forms a solid  around 1/4 
and 3/4 fillings.

The sample used is a high quality 2DES grown by molecular beam 
epitaxy with mobility, $\mu=2.4 \times 10^{7} \ {\rm cm^{2}\ V^{-1}\ 
s^{-1}} $ and density, $n=3.2 \times 10^{11}\ {\rm cm^{-2}}$.  The 
electrons are confined in a 300 ${\rm \AA}$ quantum well, 
approximately 2000 
${\rm \AA}$ beneath the surface.  A metal film coplanar waveguide 
(CPW)\cite{engel} was evaporated onto the surface of the samples.  
The length, $l$, of the CPW is 2 mm and the slot width, $w$, is 20 
$\mu$m.  The overall geometry  of the CPW is such that the line 
impedance Z$_{0} =50\ \Omega$ in the absence of the 2DES.  The real 
conductivity, 
Re[$\sigma_{xx}$], is related to the transmitted power $P$ by, 
Re[$\sigma_{xx}$] $=-\frac{w}{2 l{\rm Z}_{0}}\ln |P/P_0|$, where 
$P_0$ is the power that would be transmitted for $\sigma_{xx}=0$.  
All the data presented
 here were measured using -80 dBm inserted at the  top 
of the cryostat, although only a small fraction of this power is 
absorbed by the 2DES.  Microwave signals propagating along the CPW 
couple capacitively to the 2DES.  The microwave electric field, 
$E_m$, in the CPW is polarized perpendicular to the propagation 
direction and is mainly sensitive to the 2DES in the slot.  Two 
microwave samples were 
patterned from this wafer: one in which $E_m$ lies along the $\langle 
110 \rangle$ crystal axis or easy direction at $\nu=\frac{9}{2}$ and 
the other oriented at 90 degrees to the first so that $E_m$ lies 
along the $\langle 1 \bar1 0 \rangle$ or 
hard direction.  The upper panel of Figure 1 shows a schematic of the 
measurement circuit.  The dark 
regions are the metallic gates.  A third sample was cut adjacent to 
the other two and features 8 diffused indium contacts at the corners 
and the center of the edges.  Standard DC measurements at $T \sim 50$ 
mK on this piece show strong anisotropic transport at 
$\nu=\frac{9}{2}$ and $\frac{11}{2}$ and a well formed RIQHE.  All 
temperatures, $T$, were measured by a thermometer heat sunk directly 
to the same metal block as the sample.  

\begin{figure}[tb!]
\includegraphics[width=8cm]{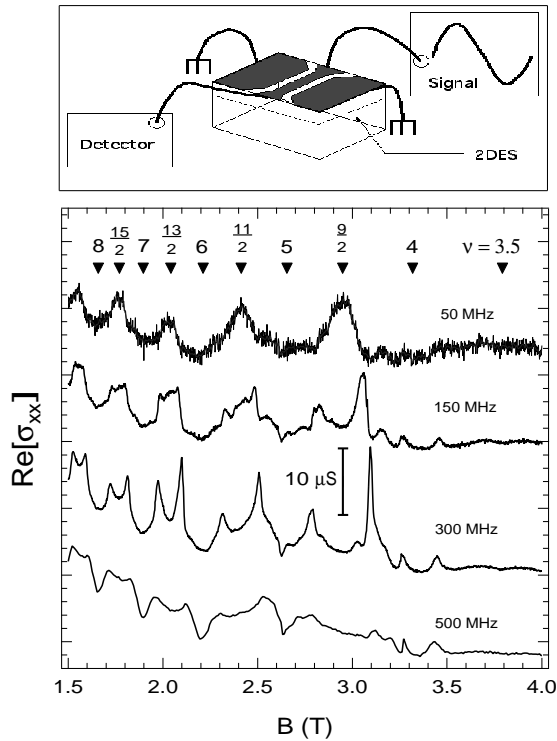}
\caption{{\bf upper panel:} A schematic of the measurement circuit.  
The dark regions represent the metallic gates. {\bf lower panel:} The real part of the 
diagonal conductivity, Re[$\sigma_{xx}$] versus $B$, magnetic field, at 50, 150, 300, and 
500 MHz.  Filling factors are marked. The data were acquired at 56 mK with the microwave electric field, $E_m$, polarized along the $\langle 110 \rangle$ {\it easy} direction.}
\label{fig.1}
\end{figure}

In lower panel of Figure 1, we plot Re[$\sigma_{xx}$] versus 
$B$ at 50, 150, 300, and 500 MHz.  The four traces are offset for 
clarity and were acquired at 56 mK.   Filling factors are marked with 
down triangles and  a scale is given in the center of the figure.  Note 
that $E_m$ is along the easy direction.  The data show well defined minima at 
integer $\nu=$8,7,6,5, and 4 and a clear dip at $\nu=3.5$.  At 50 
MHz, the transitions between the IQHE minima show a smooth increase in 
$\sigma_{xx}$ as the center of a Landau level is approached.  
However, by 150 MHz, large peaks have developed at $\nu^* 
\approx$ 1/4 and 3/4 for $\nu>4$.  These are most prominent at 300 
MHz where the peak at $\nu \approx 4.25$ is 
about 14 $\mu$S in height.  In comparison, the $\nu=4$ feature shows 
a change in $\sigma_{xx}$ of about 2 $\mu$S.  But these strong peaks 
have almost vanished in the 500 MHz data indicating that a resonance 
exists in $\sigma_{xx}$ at $\nu^*=$ 1/4 and 3/4 for $\nu>4$.  

The data in Figure 1, allow us to piece together a rough picture of the frequency dependence Re[$\sigma_{xx}$] at $\nu^*\sim$ 1/2 for $E_m$ along the easy direction.  Focusing on $\nu \sim \frac{9}{2}$, 
Re[$\sigma_{xx}$] decreases from a local maximum at 50 MHz, to a broad minimum at higher 
frequencies.  The $\nu=\frac{9}{2}$ peak drops at least 8 $\mu$S from 50 MHz to 300 MHz 
but appears to stop changing as frequency is increased further.  This 
trend is also followed by Re[$\sigma_{xx}$] at $\nu \sim \frac{11}{2}$ 
and $ \frac{13}{2}$.  Data for Re[$\sigma_{xx}$] with $E_m$ polarized along the hard direction, yield similar behavior at $f \sim$ 50 MHz but appear to show anisotropy for frequencies from 200 to 1000 MHz.
Further study of the frequency dependence about $\nu^* \sim 1/2$ is 
needed. The remainder of this paper focuses on the resonance seen in  
Re[$\sigma_{xx}$] at $\nu^* \sim$ 1/4 and 3/4.

\begin{figure}[tb!]
\begin{center}
\includegraphics[width=8cm]{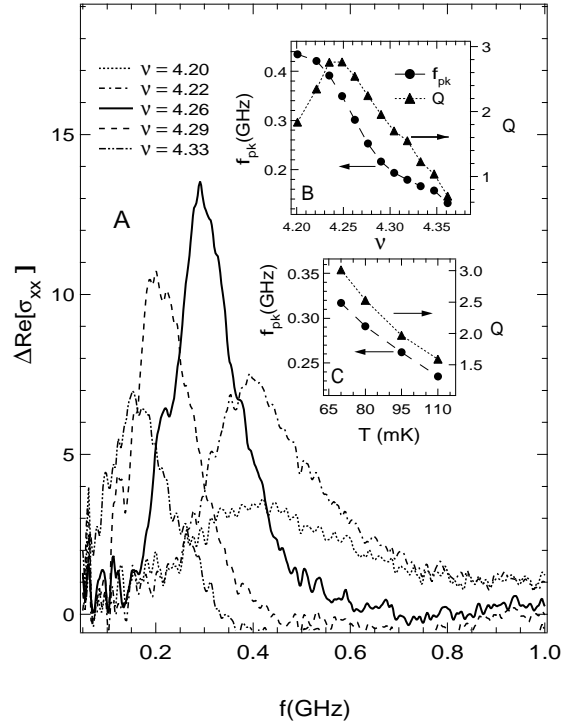}
\caption{{\bf (A)} $\Delta$Re[$\sigma_{xx}$] vs $f$ for several B 
fields in the vicinity $\nu=4.25$. $E_m$ is polarized along the {\it easy} direction. 
{\bf (B)} The peak frequency, $f_{pk}$ vs $\nu$ on the left  and Q versus $\nu$ on the right.
{\bf (C)} $f_{pk}$ and $Q $ vs $T$  at $\nu=4.26$.} 
\label{fig.2}
\end{center}
\end{figure}

The frequency dependence of $\Delta$Re[$\sigma_{xx}$]=Re[$\sigma_{xx}$]-Re[$\sigma_{bg}$] is measured with $\nu$ fixed near $4.25$ by sweeping $f$ from 50 MHz to 1 GHz.  The subtraction of a background, $\sigma_{bg}$, is necessary because the frequency response of the CPW and coaxial cables is not 
flat, although it is independent of $B$.  We measure $\sigma_{bg}$ at 
at $B$ well outside $\nu^*\sim 1/4$ and 3/4 peak structures in Figure 1 which shows flat frequency dependence with respect to $\nu=3$.
 
In Figure 2A we show the resonance in the 
$\Delta$Re[$\sigma_{xx}$] versus $f$ at  several $\nu$ from  4.20 to 
4.33.  $T \sim 56$ mK and
$E_m$ is along the easy direction.  A weak peak is first visible at 
$\nu=4.20$ (dotted line) 
where the peak frequency, $f_{pk}$, is 434 MHz. The strongest peak 
(solid line) occurs at $\nu=4.26$ where $f_{pk}=301$ MHz and the 
conductance reaches nearly 14 $\mu$S.   As filling factor is further 
increased to $\nu=4.33$, the resonance shrinks and 
$f_{pk}$ is now only 166 MHz.  
Figure 2B plots $f_{pk}$ versus $\nu$ on the left axis and $Q$ versus $\nu$
 on the right for $\nu$ from 4.20 to 4.36.  $f_{pk}$ decreases monotonically as 
the center of the level is approached. Over the same range, Q shows a 
maximum at $\nu \sim $4.25.  The range of $\nu$ of these data also shows 
that the resonance coincides with the RIQHE 
observed in Refs. \cite{mlilly1} and \cite{ rrdu}.
Figure 2C shows the  evolution of $f_{pk}$ and Q with $T$ from 70 mK 
up to 110 mK for the resonance at $\nu=4.26$.  Q and $f_{pk}$ both 
decrease as the temperature rises.  The resonance disappears at 
temperatures larger than 110 mK. 

\begin{figure}[tb]
\begin{center}
\includegraphics[width=7.5cm]{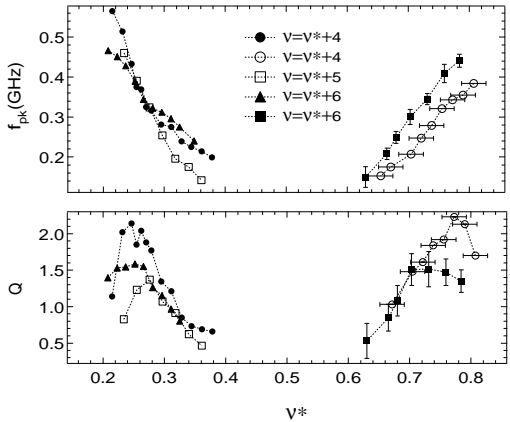}
\caption{ {\bf Upper panel} $f_{pk}$ vs $\nu^{*}$ measured for  
resonances around $\nu =$ 4.25, 4.75, 5.25, 6.25, and 6.75.  Filled symbols indicate 
measurements with $E_m$ polarized along the {\it hard} direction.  Open symbols indicate $E_m$ along the {\it easy} 
direction.  All measurements were performed at $T=56$ mK except those around $\nu=6.75$ (82 mK).
{\bf Lower panel}, $Q$ vs $\nu^{*}$ from the same set of 
resonances.   The legend is for both panels.}
\label{fig.3}
\end{center}
\end{figure}

Figure 3 presents a summary of $f_{pk}$ versus $\nu^{*}$ (upper panel) 
and $Q$ versus $\nu^*$ (lower panel) for data taken 
around $\nu=4.25$, 4.75, 5.25, 6.25 and 6.75 resonances.  Data from 
both polarizations of $E_m$ are 
shown with the closed symbols  representing measurements with $E_m$ 
along the hard direction and open symbols with $E_m$ along the 
easy.   The majority of the data were recorded at $T\approx 56$ mK 
except the $\nu \sim 6.75$ data which were taken at $T\approx 82$ 
mK.  The error in $\nu^*$ is $\pm$ 0.02 in both upper and lower 
panels and is shown on the $\nu \sim 4.75$ data.  To ensure the consistency of $\nu$ 
throughout, the magnet was always stepped down between data points.  Error bars for $f_{pk}$ and $Q$ are shown on the $\nu \sim 6.75$ data and are representative of the errors for all the data in Figure 3.

In the upper panel of Figure 3, $f_{pk}$ decreases as 
the center, $\nu^*=0.5$, of a Landau level, LL, is approached from either above or 
below.  Around $\nu^* \sim 1/4$ the measured $f_{pk}$ fall in a range 
from about 500 MHz near $\nu^*=0.20$ to about 150 MHz near 
$\nu^*=0.38$.  Around $\nu^* \sim 3/4$,  $f_{pk}$  is within the same range of 
frequencies, with $f_{pk}=150$ MHz at $\nu^*=0.63$ and $f_{pk}=441$ MHz at $\nu^*=0.78$.  
Within experimental error, there is remarkable similarity of $f_{pk}$ at different $N$, where $N= ([\nu]-2)/2$ is the LL index. Finally, the mirror image behavior of $f_{pk}$ about the LL center is consistent with particle--hole symmetry.

The data shown in Figure 3 for $\nu \sim$ 4.25 where $E_m$ is along the hard direction can be directly compared with those in Figure 2B where $E_m$ is along the easy direction.  Within error, these two data  sets for $f_{pk}$ agree from $\nu^*$=4.24 to 4.35.   This observation fits the general trend of the $f_{pk}$ data in Figure 3 which do not show a dependence on the polarization of $E_m$.  Overall,  the resonance appears to be isotropic. 

The lower panel of Figure 3 plots $Q$ versus $\nu$ for different LL and orientation of $E_m$.  The 
separate measurements agree on the range of $\nu^*$ where resonances 
are seen.  The data show the resonance is most developed for 
$\nu^*$ from 0.24 to 0.28 and $\nu^*$ from 0.71 to  0.77  where the maximum $Q$'s occur.  The smallest $Q$'s occur near the center of the LL where $Q\sim 0.5$ is seen at $\nu^*=0.36$ and 0.63.  The $Q$ of the resonance is a sharp function of $\nu^*$ with the maximally developed resonance occurring over a narrow range or $\nu^*$.  All the data show $Q$ decreasing monotonically as the center of the LL is approached.   Again, we note agreement between different filling factors and polarizations of $E_m$ and symmetry about the center of the LL.

The natural interpretation of these data is that the resonance is due to 
a pinning mode of the bubble phases around 1/4 and 3/4 filled Landau 
levels.  The low observed $f_{pk} \le $ 550 MHz means the energy of the mode is $ h \nu /k_B \le26$mK.  But, an electron oscillating in a potential well that shallow would be ionized at $T >50$ mK.  Further, $Q $ as high as 3 are measured.  The collective motion of a large region of electrons, as in a WC domain, would average the disorder and allow high $Q$.  Finally, the data are qualitatively similar to the pinning resonance observed in the high $B$ WC phase\cite{lloydssc,ccliprl97,ccliprb}.

The smooth change in $f_{pk}$ with $\nu^*$ can be explained in the 
context of a pinned electron solid in which the density is steadily 
being changed.  Increasing $\nu$ is equivalent 
to changing the density, $n^*$, of electrons that form the solid, $n^*=n \nu^*/\nu$.  In 
a weak pinning model\cite{fl1,normand}, increasing $n^*$ stiffens the 
WC domains and effectively softens the electron--disorder 
interactions which provide the restoring force\cite{ccliprb}.  Thus, around $\nu^* \sim$1/4,  
increasing $\nu^*$ results in more electrons in the bubble phase and lower 
$f_{pk}$.  Around $\nu^{*}  \sim$3/4, increasing $\nu^*$ means a bubble phase of 
holes becomes more dilute, softening the interbubble forces in 
relation to disorder, so that increasing $f_{pk}$ is observed. 

A comparison of the integrated oscillator strength, $S$, scaled by 
$f_{pk}$ for the resonance shown in Figure 2 agrees within a factor 
of 2 with the oscillator model of Fukuyama and Lee\cite{fl1,mmfoglernote}, $S/f_{pk}= n^*e\pi/2B$.   
For example, $n^*$ runs from 1.6 to 2.7 $\times \ 10^{10}$ cm$^{-2}$ where the resonance is 
seen around $\nu \sim 4.25$.  At $n^*=2.0 \times \ 10^{10}$ 
cm$^{-2}$,  we measure $S/f_{pk}=8 \times \ 10^{-6} \ \mu$S and the 
model above gives  $1.6 \times \ 10^{-5} \ \mu$S.  This supports the 
idea that only the electrons in the uppermost LL participate in the 
bubbles and the identification of the resonances as the pinning mode 
of the bubble phase.  However, where the resonance is less well 
developed, i.e. smaller Q and lower peak conductivity, the measured  
$S/f_{pk}$ is less than 1/3 of the Fukuyama and Lee model.  We note that integrating the experimental resonance data {\em underestimates} the oscillator strength because of the finite frequency range.

Using the density matrix renormalization group Shibata and 
Yoshioka\cite{yoshioka} have predicted two--electron bubbles 
exist for 0.24 $\le \nu^* \le$ 0.38 for $4< \nu < 6$.   At fillings, $6 
< \nu <8 $, they predict 2 electron bubbles from $0.18 \le \nu^* \le 
0.25$ and 3 electron bubbles for 0.29$ \le \nu^* \le$ 0.34 followed 
by a transition to a stripe phase.  Hartree--Fock calculations by the 
same authors give slightly wider ranges but predict discrete 2 and 3 
electron bubbles when $4< \nu < 6$ and 2, 3, and 4 electron bubbles 
for  $6 < \nu <8 $.  Reference \cite{foglerkoulakov96} also uses 
Hartree--Fock to predict changes in the number of electrons per bubble and other numerical work\cite{kunyang} predicts bubbles at $\nu^*=1/3$ and 1/4.  Our data 
for $Q$ in Figure 2B and Figure 3 is in rough agreement with the predicted range 
for bubbles.  However, our data give
neither a way to know the number of electrons per bubble nor a reason 
to suspect a change in that number within a given data set.

In summary, we have observed a sharp resonance in the real part of the 
microwave conductivity at $\nu^{*}=1/4$ and 3/4 starting at  $\nu=4.25$ and going at least as 
high as $\nu=6.75$.  Changing the orientation of the microwave 
electric field with respect to the GaAs lattice has only minor 
effects on the resonance.  Above $T \sim 110$ mK the resonance
fades. The resonance is visible for a range of $\nu^*$ from 0.20 to 0.38
 and again from 0.64 to 0.80.  These ranges of $\nu^*$ coincide with observations
 of the RIQHE and theoretical discussions of the bubble 
phase.  The data presented offer the most direct evidence to date 
that the re--entrant insulating behavior of the 2DES around $\nu^* \sim 1/4$ and 3/4 in higher LL  is caused by the formation of a crystalline bubble phase.

This work is supported by the AFOSR and the NHMFL in--house research program.

\end{document}